\documentclass[12pt,a4paper]{article}

\usepackage{bm}

\usepackage{amsmath}
\usepackage{amsfonts}
\usepackage{amssymb}

\usepackage[dvipdfmx]{graphicx}

\usepackage{colortbl}
\usepackage{hhline}

\newcommand{\CC}{\mathbb{C}}
\newcommand{\ZZ}{\mathbb{Z}}

\newcommand{\ol}{\overline}
\newcommand{\wt}{\widetilde}

\def\Res{\mathop{\rm Res}}
\def\tr{\mathop{\rm tr}\nolimits}
\def\Pexp{\mathop{\rm Pexp}\nolimits}

\begin{document}
\begin{titlepage}
\title{
\vspace{-1.5cm}
\begin{flushright}
{\normalsize TIT/HEP-686\\ August 2021}
\end{flushright}
\vspace{1.5cm}
\LARGE{Finite-$N$ superconformal index via the AdS/CFT correspondence}}
\author{
Yosuke {\scshape Imamura\footnote{E-mail: imamura@phys.titech.ac.jp}}
\\
\\
{\itshape Department of Physics, Tokyo Institute of Technology}, \\ {\itshape Tokyo 152-8551, Japan}}

\date{}
%\date{July 12, 2019}
\maketitle
\thispagestyle{empty}
\begin{abstract}
We propose a prescription to calculate the superconformal index of the ${\cal N}=4$ $U(N)$
supersymmetric Yang-Mills theory with finite $N$
on the AdS side.
The finite $N$ corrections are included as contributions of D3-branes wrapped around three-cycles in ${\bm S}^5$,
which are calculated as the index of the gauge theories
realized on the wrapped branes.
The single-wrapping contribution has been studied in a previous work,
and we further confirm that the inclusion of multiple-wrapping contributions
correctly reproduces the higher order terms
as far as we have checked numerically.
\end{abstract}
\end{titlepage}

\tableofcontents
%%%%%%%%%%%%%%%%%%%%%%%%%%%%%%%%%%%%%%%%%%%%%%%%%%%%%%%%%%%%%%%%%%%%%%%%%%%%%%%%
\section{Introduction}
The AdS/CFT correspondence \cite{Maldacena:1997re}
has been intensively investigated in the large $N$ limit,
and agreement of many quantities calculated on the both side of the duality has been confirmed.
Although the quantum gravitational effect becomes important for finite $N$,
it may be possible to calculate some quantities associated with topology and supersymmetry
on the AdS side.
Let us consider the duality between 4d ${\cal N}=4$ SYM and the string theory in $AdS_5\times S^5$.
The relation $L^4/l_p^4\sim T_{\rm D3}L^4\sim N$ among
the AdS radius $L$, the ten-dimensional Planck length $l_p$,
and the D3-brane tension $T_{\rm D3}$
suggests that when $N$ is finite not only the quantum gravitational effect
but also the contribution of D3-branes extended in $AdS_5\times S^5$ becomes important,
and the finite $N$ corrections for some quantities may be reproduced as the
contribution of D3-branes.
One such example is the BPS partition function of ${\cal N}=4$ SYM.
It was calculated by the geometric quantization of BPS D3-brane configurations (sphere giants
\cite{McGreevy:2000cw,Mikhailov:2000ya})
in $S^5$ \cite{Biswas:2006tj}.
(See also \cite{Mandal:2006tk} for another derivation using AdS giants
\cite{Grisaru:2000zn,Hashimoto:2000zp}.)
In this paper we discuss a similar calculation of the superconformal index \cite{Kinney:2005ej}
for finite $N$ on the AdS side.

The basic idea is as follows.
We start from the large $N$ limit, in which the index is reproduced as the supergravity contribution in the holographic dual description \cite{Kinney:2005ej}.
We include finite $N$ corrections as contributions of
branes wrapped around topologically-trivial three-cycles in the internal space $S^5$.
The analysis of single-wrapping contributions
has been already done in \cite{Arai:2019xmp},
and it was confirmed that finite $N$ corrections are correctly reproduced
up to errors which can be interpreted as the multiple-wrapping contributions.
The multiple-wrapping contributions were calculated
in  \cite{Arai:2020qaj} for the Schur index \cite{Gadde:2011uv} of 4d ${\cal N}=4$ SYM,
and the analytic result of \cite{Bourdier:2015wda} was successfully reproduced.
In this paper we improve the calculation in \cite{Arai:2020qaj},
and confirm that we can also calculate the
superconformal index in a similar way.

In terms of ${\cal N}=1$ multiplets
the ${\cal N}=4$ theory consists of a $U(N)$ vector multiplet
and three adjoint chiral multiplets $\Phi_I$ ($I=1,2,3$).
The superconformal index
is defined by
\begin{align}
{\cal I}=\tr_{\rm BPS}[
e^{2\pi i(J+\ol J)}
q^{H+\ol J}
y^{2J}
u_1^{R_1}
u_2^{R_2}
u_3^{R_3}
],\quad
u_1u_2u_3=1.
\label{scindex}
\end{align}
We use the Hamiltonian $H$, left- and right-handed spins $J$ and $\ol J$,
and R-charges $R_I$ as the Cartan generators of the superconformal algebra $psu(2,2|4)$.
$R_I$ acts on the corresponding chiral field $\Phi_I$ with charge $1$.
\footnote{We use unusual normalization of R-charges such that $\cal Q$ carries $R_I=1/2$.}
The supercharge ${\cal Q}$ associated with the index (\ref{scindex}) carries the following quantum numbers:
\begin{align}
{\cal Q}:(H,J,\ol J,R_1,R_2,R_3)=(+\tfrac{1}{2},0,-\tfrac{1}{2},+\tfrac{1}{2},+\tfrac{1}{2},+\tfrac{1}{2}).
\end{align}

\begin{figure}[htb]
\centering
\includegraphics{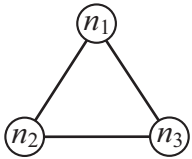}
\caption{The quiver diagram of the theory realized on the D3-brane system with the wrapping numbers $(n_1,n_2,n_3)$.}\label{c3quiver.eps}
\end{figure}

The formula we use to calculate the index has the following form:
\begin{align}
{\cal I}_{U(N)}={\cal I}_{\rm KK}\sum_{n_1,n_2,n_3=0}^\infty{\cal I}_{(n_1,n_2,n_3)},
\label{theformula}
\end{align}
where ${\cal I}_{\rm KK}$ is the supergravity contribution
giving the large $N$ index and the sum over $n_I$ gives the finite $N$
corrections arising from wrapped D3-branes.
$n_I$ are the numbers of D3-branes wrapped on three-cycles
$S_I\subset S^5$ labeled by $I=1,2,3$.
Let $z_I$ be the three complex coordinates
corresponding to $\Phi_I$.
The internal space $S^5$ is given by $|z_1|^2+|z_2|^2+|z_3|^2=1$.
Each of $R_I$ generates phase rotation of the corresponding $z_I$-plane.
Three three-cycles $S_I$ ($I=1,2,3$) are defined by $z_I=0$.
For a set of the wrapping numbers $(n_1,n_2,n_3)$,
the gauge theory realized on the wrapped D3-branes
is the $U(n_1)\times U(n_2)\times U(n_3)$ gauge theory
with three bi-fundamental hypermultiplets
shown as the quiver diagram in Figure \ref{c3quiver.eps}.
This looks like the toric diagram of $\CC^3$.
This is not accidental but holds for general toric quiver gauge theories \cite{Arai:2019aou}.
${\cal I}_{(n_1,n_2,n_3)}$ is the index of the D-brane system specified by the wrapping numbers
and ${\cal I}_{(0,0,0)}=1$.

To sort out the results of our calculation it is convenient to define the
total wrapping number
\begin{align}
n=n_1+n_2+n_3.
\end{align}
The contribution from each $n$ is of order
$q^{nN+\delta_n}$ where $\delta_n\geq0$ are
non-negative integers.
In the large $N$ limit the sectors with $n\geq1$ decouples and
only the $n=0$ sector contributes to the index.
Then (\ref{theformula}) reduces to ${\cal I}_{U(\infty)}={\cal I}_{\rm KK}$,
which was confirmed in \cite{Kinney:2005ej}.

The contribution of the $n=1$ sector
was investigated in \cite{Arai:2019xmp}.
In this case the theory on the wrapped D3-brane is a
supersymmetric $U(1)$ gauge theory consisting of only neutral fields, and
we can easily calculate ${\cal I}_{(n_1,n_2,n_3)}$ without any holonomy integrals.
It was found that
(\ref{theformula}) reproduces the correct index
up to expected error due to multiple-wrapping contribution of order $q^{2N+\delta_2}$ with $\delta_2=4$.

The purpose of this paper is to study the contributions from $n\geq2$.
There are two issues we have to settle for the purpose.
One is about the gauge fugacity integrals.
If $n\geq2$, the theory on wrapped D3-branes is a gauge theory
with charged fields.
We can write down the contribution in the matrix integral form
just like standard localization formulas.
In order to carry out the gauge fugacity integrals we need to carefully specify the integration contours,
because
due to a reason we will explain in Section \ref{problem.sec} we cannot use the standard choice,
the unit circle on the complex plane of the gauge fugacities.
The other is related to a global symmetry of the quiver gauge theory of Figure \ref{c3quiver.eps}.
A gauge invariant operator of this theory
can be associated with a closed path in the quiver diagram.
If all $n_I$ are positive,
there exist operators corresponding to
paths going around the triangle.
Let $U(1)_{\rm loop}$ be the global symmetry
coupling to such operators.
Because there is no corresponding symmetry in the boundary SYM, the $U(1)_{\rm loop}$
fugacity variable $a_{\rm loop}$
cannot be an independent variable
but it must be given as a function of other fugacities.
We have to fix these two ambiguities, one for integration contours and the other for $a_{\rm loop}$.
We will
propose how to fix these
and confirm it gives the correct index.

This paper is organized as follows.
In the next section we gives a detailed description of the formula
(\ref{theformula}) except for the ambiguities mentioned above.
In Section \ref{problem.sec}
we propose a basic rule to determine the integration contours.
In fact, the ambiguity for the integration contours
is not fixed only by the rule in Section \ref{problem.sec}.
The ambiguities concerning the contours and $a_{\rm loop}$
are fixed in section \ref{s5.sec} by using
``the pole cancellation condition,''
which requires the index does not diverge
in the unrefined limit $u_I\rightarrow 1$.
We also numerically confirm in Section \ref{s5.sec} that
the prescription reproduces the correct finite $N$ index
up to very high order of $q$.
The final section will be devoted to summary and discussion.

%%%%%%%%%%%%%%%%%%%%%%%%%%%%%%%%%%%%%%%%%%%%%%%%%%%%%%%%%%%%%%%%%
\section{Formula}
In this section we give a detailed explanation of the formula (\ref{theformula}).
The basic idea was first given in \cite{Arai:2019xmp}.

In the large $N$ limit it is well known that the index
is reproduced on the AdS side as the supergravity index
${\cal I}_{\rm KK}=\Pexp i_{\rm KK}$ \cite{Kinney:2005ej}.
$\Pexp$ is the plethystic exponential defined in the next section.
$i_{\rm KK}$ is the single-particle index
\begin{align}
i_{\rm KK}
&=\frac{qu_1}{1-qu_1}
+\frac{qu_2}{1-qu_2}
+\frac{qu_3}{1-qu_3}
-\frac{q^{\frac{3}{2}}y}{1-q^{\frac{3}{2}}y}
-\frac{q^{\frac{3}{2}}y^{-1}}{1-q^{\frac{3}{2}}y^{-1}}.
\end{align}

The finite $N$ corrections
${\cal I}_{(n_1,n_2,n_3)}$
in (\ref{theformula})
are given by
\begin{align}
{\cal I}_{(n_1,n_2,n_3)}
=
{\cal I}_{\rm cl}
\int_C d\mu_1 d\mu_2 d\mu_3
(\mbox{vector})
(\mbox{hyper}).
\label{in1n2n3}
\end{align}
${\cal I}_{\rm cl}$ is the classical factor coming from the classical charges and energy of the wrapped brane system.
Its explicit form is
\begin{align}
{\cal I}_{\rm cl}
=(qu_1)^{n_1N}
(qu_2)^{n_2N}
(qu_3)^{n_3N},
\label{iclassical}
\end{align}
and this is the only place in the formula where the rank $N$ appears.
$d\mu_I$ are the integration measures associated with $U(n_I)$ gauge group.
\begin{align}
d\mu_I=\frac{1}{n_I!}\prod_{a=1}^{n_I}\frac{dz_{I,a}}{2\pi iz_{I,a}}
\prod_{a\neq b}\left(1-\frac{z_{I,a}}{z_{I,b}}\right).
\label{dmeasure}
\end{align}
$z_{I,a}$ ($a=1,2,\ldots,n_I$)
are gauge fugacities of $U(n_I)$.

The integrand consists of two factors.
``(vector)'' is the contribution from the vector multiplets
and given by
\begin{align}
(\mbox{vector})
=\prod_{I=1}^3
\prod_{a=1}^{n_I}
\prod_{b=1}^{n_I}
\Pexp
f_v^I\left(\frac{z_{I,a}}{z_{I,b}}-\frac{z_{I,b}}{z_{I,a}}\right),
\end{align}
where $f_v^I$ are the single-particle indices of $U(1)$ vector multiplets.
$f_v^3$ is given by
\begin{align}
f_v^3
&=1-\frac{(1-q^{-1}u_3^{-1})
(1-q^{\frac{3}{2}}y)(1-q^{\frac{3}{2}}y^{-1})
}{(1-qu_1)(1-qu_2)}.
\label{is1}
\end{align}
$f_v^1$ and $f_v^2$ are obtained by permutations among $u_I$.

The hypermultiplet contribution ``(hyper)'' is
\begin{align}
(\mbox{hyper})
=\prod_{I=1}^3
\prod_{a=1}^{n_I}
\prod_{b=1}^{n_{I+1}}
\Pexp
f^{I,I+1}_h\left(a_{I,I+1}\frac{z_{I,a}}{z_{I+1,b}}
-a_{I,I+1}^{-1}\frac{z_{I+1,b}}{z_{I,a}}\right),
\label{hyper1}
\end{align}
where $I$ is treated as a cyclic variable and when $I=3$ $I+1$ means $1$.
Three variables $a_{12}$, $a_{23}$, and $a_{31}$ are (redundant) $U(1)_{\rm loop}$ fugacities,
and the integral depends only on their product $a_{\rm loop}=a_{12}a_{23}a_{31}$.
If some of $n_I$ are zero, we can absorb
these degrees of freedom by a redefinition of
gauge fugacities and
we can neglect them.
In the boundary gauge theory there is no parameter corresponding to $a_{\rm loop}$,
and the associated symmetry should  be broken.
This means $a_{\rm loop}$ is not an independent variable but a function of other fugacities.
$f_h^{IJ}$ are the single-particle indices of
hypermultiplets.
$f_h^{12}$ is given by
\begin{align}
f_h^{12}
&=\frac{u_3^{\frac{1}{2}}}{q}\frac{(1-q^{\frac{3}{2}}y)(1-q^{\frac{3}{2}}y^{-1})}{1-qu_3}.
\label{is12}
\end{align}
$f_h^{23}$ and $f_h^{31}$ are obtained from this by the permutations.
For derivations of (\ref{is1}) and (\ref{is12}) see Appendix \ref{spi.app}.

%%%%%%%%%%%%%%%%%%%%%%%%%%%%%%%%%%%%%%%%%%%%%%%%%%%%%%%%%%%%%%%%%
\section{Pole selection rule}\label{problem.sec}
The contribution of a single D3-brane wrapping around a cycle $S_I$
is given by $(qu_I)^N\Pexp i_v^I$.
The plethystic exponential $\Pexp i$ is defined as follows.
The $q$-expansion of $i$ in general has the form
\begin{align}
i=\sum_rf_r-\sum_sg_s,
\label{iexpa}
\end{align}
where the first sum and the second sum are bosonic and fermionic contributions, respectively.
Each term $f_r$ or $g_s$ gives a quantum of a harmonic oscillator.
Summing up the contribution from all states of the oscillator
we obtain
\begin{align}
\Pexp(f_r)=1+f_r+f_r^2+\cdots,
\label{bosonicsum}
\end{align}
for a bosonic oscillator and
\begin{align}
\Pexp(-g_s)=1-g_s,
\label{fermionicsum}
\end{align}
for a fermionic oscillator.
The plethystic exponential
$\Pexp$ is
defined by these two relations and the addition formula
$\Pexp (f+g) = (\Pexp f)(\Pexp g)$
representing the decoupling among oscillators.

An unusual point in the calculation of the wrapped D-brane contribution
is the existence of tachyonic modes, which have negative energies.
Such modes correspond to terms
in $i$ with negative power of $q$.
The existence of
such modes are related to the fact that the cycle
is topologically trivial.
Usually we take $|q|<1$ and other fugacities to be phases,
and if there exists such tachyonic modes
the geometric series in
(\ref{bosonicsum}) does not converge.
In \cite{Arai:2019xmp} it was proposed that
when the geometric series
(\ref{bosonicsum})
diverges we should define the plethystic exponential by
analytically continued form:
\begin{align}
\Pexp(f_r)=\frac{1}{1-f_r}.
\label{analyticcontinuation}
\end{align}
Namely, we define the plethystic exponential
of (\ref{iexpa}) by
\begin{align}
\Pexp i=\frac{\prod_s(1-g_s)}{\prod_r(1-f_r)}.
\label{pexpdef}
\end{align}
As far as we have checked
this trick works quite well and correct indices are reproduced in many cases.
If $f_r$ includes negative power of $q$
the $q$-expansion of its plethystic exponential around $q=0$
starts at positive power of $q$.
This is the origin of the tachyonic shift $\delta_n>0$
for $n\geq1$.

In the case of $n=2$, the gauge group realized on the wrapped brane system
is $U(1)^2$ or $U(2)$.
Because the diagonal $U(1)$ always decouples
we have one non-trivial gauge integral.
It typically has the form
\begin{align}
\oint_C\frac{dz}{2\pi iz}\Pexp i(z+z^{-1}),
\label{rank1}
\end{align}
where $i$ is a $q$-series of the form (\ref{iexpa}).
The integrand has poles at $z=f_r$ and $z=f_r^{-1}$
for each $r$, and these two poles have opposite residues.
Let us call these two series of poles ``positive poles'' and ``negative poles'', respectively.
The problem is which poles we should pick up in the integral (\ref{rank1}).

If we consider $n>2$ case the situation becomes more complicated.
We need to consider the multiple integrals.
Let $z_1,z_2,\ldots,z_n$ be the gauge fugacities,
and let us carry out the integrals in this order.
Because all fields belong to the adjoint or bi-fundamental representations,
we can set the last one $z_n$ to be one.
Let us consider the $i$-th integral ($1\leq i\leq n-1$).
The gauge fugacity $z_i$ appears in the single-particle index in the form
$f_r\frac{z_k}{z_i}$ or
$f_r\frac{z_i}{z_k}$.
These produce the poles at
$z_i=f_rz_k$ and $z_i=f_r^{-1}z_k$, respectively.
If $k>i$, the $z_k$ integral has not yet been performed
at the moment of the $z_i$ integral,
and we treat the former and the latter as a positive pole and a negative pole,
respectively.
However, if $k<i$, $z_k$ has been replaced with the position of a pole on the $z_k$ plane
by the $z_k$ integral.
By taking account of this, the position of a pole
on the $z_i$ plane in general has the following form
\begin{align}
z_i=\frac
{f_{r_1}f_{r_2}\cdots f_{r_p}}{f_{r'_1}f_{r'_2}\cdots f_{r'_q}}z_j,
\end{align}
where $j>i$.
There are three types of poles
\begin{itemize}
\item $p\geq1$, $q=0$ : positive poles
\item $p=0$, $q\geq1$ : negative poles
\item $p\geq1$, $q\geq1$ : mixed poles
\end{itemize}
Which poles should we pick up in the $z_i$-integral?
In the usual situation with $|f_r|<1$ and
the unit circle as the integration contour,
the contour encloses all positive poles (and $z=0$).
Naively, the contour seems to enclose
a part of mixed poles.
In fact, however,
we can show that the mixed poles do not appear
due to a certain cancellation (See Appendix \ref{mixed.sec}).
Therefore, in the standard situation
with $|f_r|<1$, using the unit circles
is equivalent to taking only positive poles at each step of
integrals.

Our proposal is to keep using this prescription
regardless of $|f_r|$
being inside or outside of the unit circle.
\begin{description}
\item[The pole selection rule]\mbox{}

In the case with the wrapping number $n$,
we have $n$ gauge fugacities $z_i$ ($i=1,2,\ldots,n$).
Let us suppose that we carry out the integral in the order $z_1,z_2,\ldots,z_n$.
Because the last $z_n$-integral is trivial,
we can set $z_n=1$ and we have only $n-1$ non-trivial integrals.
In the $z_k$-integral at the $k$-th step ($1\leq k<n$)
we have the following two types of poles
other than $z_k=0$:
\begin{align}
z_k=fz_{l(\neq k)},\quad
z_k=f^{-1}z_{l(\neq k)},
\end{align}
where the former and the latter
are associated with
terms $+fz_l/z_k$ and $+fz_k/z_l$ in the single-particle index, respectively.
The rule we propose is to include poles at $z_k=0$ and $z_k=fz_{l(\neq k)}$
and exclude ones at $z_k=f^{-1}z_{l(\neq k)}$.
When we choose a pole $z_k=fz_l$ with $l<k$ we substitute to $z_l$ the pole position
chosen in the preceding $z_l$-integral.
\end{description}
This rule does not refer to the numerical values of $f_r$,
and gives the index as an analytic function of $f_r$.

The analysis in \cite{Arai:2020qaj} of the Schur index supports this rule.
The Schur index is defined from the superconformal index by taking the limit
$y\rightarrow u_3q^{-\frac{1}{2}}$ \cite{Gadde:2011uv}.
In this limit the index becomes
a function of only two fugacities
$q'=qu_3^{-\frac{1}{2}}$ and
$u'=u_1^{\frac{1}{2}}u_2^{-\frac{1}{2}}$,
and the pole structure in the gauge fugacity integrals
is much simpler than that of the superconformal index.
In \cite{Arai:2020qaj} it was found that if we pick up positive poles
at each step of the gauge fugacity integrals
the correct Schur index is reproduced.

\newcommand{\eqo}{{\stackrel{\circ}{=}}}

%%%%%%%%%%%%%%%%%%%%%%%%%%%%%%%%%%%%%%%%%%%%%%%%%%%%%%%%%%%%%%%
\section{Comparison}\label{s5.sec}
We can calculate the index unambiguously on the gauge theory side
by using the localization formula.
The results for $N=1,2,3,4$ are
\begin{align}
{\cal I}^{\rm gauge}_{U(1)}&\eqo 1+3q-2q^{\frac{3}{2}}+3q^2+6q^{\frac{7}{2}}-6q^4+12q^5-18q^{\frac{11}{2}}+27q^6+\cdots,\nonumber\\
{\cal I}^{\rm gauge}_{U(2)}&\eqo 1+3q-2q^{\frac{3}{2}}+9q^2-6q^{\frac{5}{2}}+11q^3-6q^{\frac{7}{2}}+9q^4+14q^{\frac{9}{2}}+\cdots,\nonumber\\
{\cal I}^{\rm gauge}_{U(3)}&\eqo 1+3q-2q^{\frac{3}{2}}+9q^2-6q^{\frac{5}{2}}+21q^3-18q^{\frac{7}{2}}+33q^4-22q^{\frac{9}{2}}+\cdots,\nonumber\\
{\cal I}^{\rm gauge}_{U(4)}&\eqo 1+3q-2q^{\frac{3}{2}}+9q^2-6q^{\frac{5}{2}}+21q^3-18q^{\frac{7}{2}}+48q^4-42q^{\frac{9}{2}}+\cdots
\label{isun}
\end{align}
(We set $y=u_I=1$ and leave only $q$ to save the space.
We use the notation ``$\eqo$'' to represent this unrefined limit.)
We want to reproduce these on the gravity side.

\subsection{$n=0$}
Let us compare these with the supergravity contribution ${\cal I}_{\rm KK}$.
The $q$-expansion of ${\cal I}_{\rm KK}$ is
\begin{align}
{\cal I}_{\rm KK}&\eqo 1+3q-2q^{\frac{3}{2}}+9q^2-6q^{\frac{5}{2}}+21q^3-18q^{\frac{7}{2}}+48q^4-42q^{\frac{9}{2}}+\cdots
\end{align}
Let $\Delta{\cal I}_{U(N)}^{(1)}\equiv{\cal I}^{\rm gauge}_{U(N)}-{\cal I}_{\rm KK}$ be the
finite $N$ correction.
For small $N$ it is given by
\begin{align}
\Delta{\cal I}_{U(1)}^{(1)}&\eqo -6q^2+6q^{\frac{5}{2}}-21q^3+24q^{\frac{7}{2}}-54q^4+42q^{\frac{9}{2}}-87q^5+78q^{\frac{11}{2}}+\cdots,\nonumber\\
\Delta{\cal I}_{U(2)}^{(1)}&\eqo -10q^3+12q^{\frac{7}{2}}-39q^4+56q^{\frac{9}{2}}-120q^5+132q^{\frac{11}{2}}-217q^6+\cdots,\nonumber\\
\Delta{\cal I}_{U(3)}^{(1)}&\eqo -15q^4+20q^{\frac{9}{2}}-63q^5+102q^{\frac{11}{2}}-219q^6+288q^{\frac{13}{2}}-480q^7+\cdots,\nonumber\\
\Delta{\cal I}_{U(4)}^{(1)}&\eqo -21q^5+30q^{\frac{11}{2}}-93q^6+162q^{\frac{13}{2}}-351q^7+510q^{\frac{15}{2}}-876q^8+\cdots.
\label{error1}
\end{align}
The leading term of each correction is of order $q^{N+1}$.
We want to reproduce these corrections by wrapped D3-branes on the AdS side.

%%%%%%%%%%%%%%%%%%%%%%%%%%%%%%%%%%%%%%%%%%%%%%%%%%%%%%%%%%
\subsection{$n=1$}
The $n=1$ contribution was investigated in \cite{Arai:2019xmp}.
For $(n_1,n_2,n_3)=(0,0,1)$ the index (\ref{in1n2n3}) reduces to
\begin{align}
{\cal I}_{(0,0,1)}&=(qu_3)^N\Pexp f_v^3.
\end{align}
By using the prescription for the tachyonic mode this gives
\begin{align}
{\cal I}_{(0,0,1)}
=\frac{(qu_3)^N}{(1-\frac{u_1}{u_3})(1-\frac{u_2}{u_3})}
\left[-u_3q
+(y+y^{-1})q^{\frac{3}{2}}
+\cdots\right].
\label{is1eq23}
\end{align}
The other two single-wrapping contributions from $S_1$ and $S_2$
are obtained in a similar way.
The total contribution from $n=1$ is
\begin{align}
{\cal I}^{(1)}_{U(N)}={\cal I}_{\rm KK}({\cal I}_{(1,0,0)}+{\cal I}_{(0,1,0)}+{\cal I}_{(0,0,1)})
\label{ikkis1is2is3}
\end{align}
and for $N=1,2,3,4$ it is given by
\begin{align}
{\cal I}^{(1)}_{U(1)}
&\eqo -6q^2+6q^{\frac{5}{2}}-21q^3+24q^{\frac{7}{2}}-54q^4+42q^{\frac{9}{2}}-87q^5
+78q^{\frac{11}{2}}+\cdots,\nonumber\\
{\cal I}^{(1)}_{U(2)}
&\eqo -10q^3+12q^{\frac{7}{2}}-39q^4+56q^{\frac{9}{2}}-120q^5+132q^{\frac{11}{2}}-217q^6
+\cdots,\nonumber\\
{\cal I}^{(1)}_{U(3)}
&\eqo -15q^4+20q^{\frac{9}{2}}-63q^5+102q^{\frac{11}{2}}-219q^6
+288q^{\frac{13}{2}}-480q^7+\cdots,\nonumber\\
{\cal I}^{(1)}_{U(4)}
&\eqo -21q^5+30q^{\frac{11}{2}}-93q^6
+162q^{\frac{13}{2}}-351q^7+510q^{\frac{15}{2}}-876q^8+\cdots.
\label{in1234n1}
\end{align}
These correctly reproduces many terms appearing in (\ref{error1}).

Although we only show the results in the unrefined limit
in (\ref{in1234n1}),
we cannot set $u_I=1$ before summing up the three contributions
in (\ref{ikkis1is2is3}).
As is seen in (\ref{is1eq23})
each of ${\cal I}_{(n_1,n_2,n_3)}$ has the pole at $u_I=1$.
The most singular part in the $\epsilon_I=\log u_I$ expansion
in the leading term of the $q$-expansion is
\begin{align}
-\frac{1}{(\epsilon_3-\epsilon_1)(\epsilon_3-\epsilon_2)}q^{N+1}
\end{align}
and diverges at $\epsilon_I=0$.
The poles in ${\cal I}_{(n_1,n_2,n_3)}$ cancel only after three contributions are summed
in (\ref{ikkis1is2is3}).
This pole cancellation plays an important role
when we discuss $n\geq2$ contributions.

Even after taking account of the single-wrapping contributions,
we still have the error
\begin{align}
\Delta{\cal I}_{U(N)}^{(2)}:=
{\cal I}^{\rm gauge}_{U(N)}
-({\cal I}_{\rm KK}+{\cal I}_{U(N)}^{(1)}).
\end{align}
For $N=1,2,3,4$ the remaining errors are as follows.
\begin{align}
\Delta{\cal I}_{U(1)}^{(2)}&\eqo 112q^6-516q^{\frac{13}{2}}+1869q^7-5394q^{\frac{15}{2}}+14010q^8-32850q^{\frac{17}{2}}+\cdots,\nonumber\\
\Delta{\cal I}_{U(2)}^{(2)}&\eqo 252q^8-1218q^{\frac{17}{2}}+4477q^9-13164q^{\frac{19}{2}}+34452q^{10}+\cdots,\nonumber\\
\Delta{\cal I}_{U(3)}^{(2)}&\eqo 504q^{10}-2548q^{\frac{21}{2}}+9549q^{11}-28734q^{\frac{23}{2}}+76506q^{12}+\cdots,\nonumber\\
\Delta{\cal I}_{U(4)}^{(2)}&\eqo 924q^{12}+\cdots.
\label{singleerror}
\end{align}
The leading term for each $N$ is of order $q^{2N+4}$.
Our next task is to reproduce these corrections as double-wrapping contributions.

%%%%%%%%%%%%%%%%%%%%%%%%%%%%%%%%%%%%%%%%%%%%%%%%%%%%%%%%%%%%%%%%%%%%%%%%%%%
\subsection{$n=2$}
\begin{figure}[htb]
\centering
\includegraphics{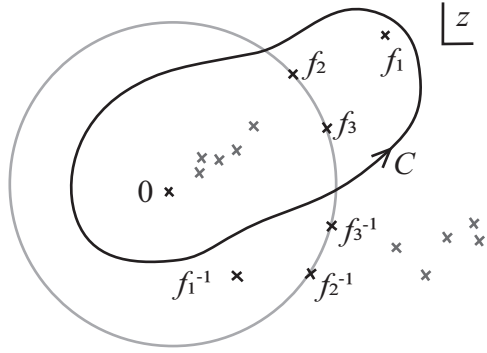}
\caption{The pole structure of the integrand in (\ref{s1u2}) is shown.}\label{wcontour.eps}
\end{figure}

There are six configurations
with total wrapping number $2$.
\begin{align}
{\cal I}_{U(N)}^{(2)}={\cal I}_{\rm KK}({\cal I}_{(2,0,0)}
+{\cal I}_{(0,2,0)}
+{\cal I}_{(0,0,2)}
+{\cal I}_{(1,1,0)}
+{\cal I}_{(0,1,1)}
+{\cal I}_{(1,0,1)})
\label{sixterms}
\end{align}
Each of the first three terms is contribution of
two D3-branes wrapping on the same cycle,
on which $U(2)$ gauge theory is realized.
The Cartan $U(1)^2$ part simply gives the square of the
single-wrapping contribution.
In addition we need to include the
W-boson contribution.
For example, ${\cal I}_{(0,0,2)}$
is given by
\begin{align}
{\cal I}_{(0,0,2)}={\cal I}_{(0,0,1)}^2{\cal I}_{W_3}
\end{align}
with the W-boson contribution
\begin{align}
{\cal I}_{W_3}
&=\frac{1}{2}
\oint_C\frac{dz}{2\pi i}\Pexp((f_v^3-1)(z+z^{-1})).
\label{s1u2}
\end{align}
Let us look at the pole structure of the integrand.
The $q$-expansion of $f_v^3-1$ is
\begin{align}
f_v^3-1=\frac{1}{qu_3}-1+\frac{u_1}{u_3}+\frac{u_2}{u_3}
-\frac{y}{u_3}q^{\frac{1}{2}}
-\frac{1}{u_3y}q^{\frac{1}{2}}
+{\cal O}(q).
\end{align}
Let $f_r$ ($r=1,2,3,\ldots$) be the positive terms in the expansion.
There are one tachyonic term and two zero-mode terms.
\begin{align}
f_1=\frac{1}{qu_3},\quad
f_2=\frac{u_1}{u_3},\quad
f_3=\frac{u_2}{u_3}.
\end{align}
In addition, we have infinite number of $f_r$ ($r\geq3$) with positive exponent of $q$.
According to the rule we take the contour enclosing all positive poles
(Figure \ref{wcontour.eps}).
The result of the contour integral is\footnote{In the practical calculation
we need to introduce a cut-off in the $q$-expansion of the single-particle index.
For a detailed explanation about the cut-off order see Appendix \ref{cutoff.app}.}
\begin{align}
{\cal I}_{W_3}
&=\frac{2u_3^2q^2+(-2u_1-2u_2+2u_3-u_3^{-2}-u_3^4)(y+y^{-1})q^{\frac{5}{2}}+\cdots}{(1+\frac{u_1}{u_3})(1+\frac{u_2}{u_3})(1-u_3^{-3})}.
\end{align}
Just like the single wrapping contributions,
${\cal I}_{(0,0,2)}$ has a pole at $u_I=1$,
which comes from both ${\cal I}_{(0,0,1)}^2$ and ${\cal I}_{W_3}$.
The most singular part in the leading term of ${\cal I}_{(0,0,2)}$ is
\begin{align}
{\cal I}_{(0,0,2)}=\left(
\frac{1}{6\epsilon_3(\epsilon_3-\epsilon_1)^2(\epsilon_3-\epsilon_2)^2}+\mbox{less singular terms}
\right)q^{2N+4}
+\cdots
\label{wbospole}
\end{align}
Notice that this pole does not cancel even in the sum
${\cal I}_{(2,0,0)}+{\cal I}_{(0,2,0)}+{\cal I}_{(0,0,2)}$.
This should be cancelled by the other three contributions including the hypermultiplets.

Each of the last three terms in (\ref{sixterms}) is
the contribution of two D3-branes wrapping on two different cycles.
The gauge group on the branes is $U(1)^2$ and fields on each brane are
neutral.
Therefore, the contribution from the vector multiplets is
simply the product of single-wrapping contributions for two branes.
The extra contribution we need to calculate
is the hypermultiplet arising from open strings
stretched between two D3-branes.
For example, ${\cal I}_{(1,1,0)}$ is given by
\begin{align}
{\cal I}_{(1,1,0)}={\cal I}_{(1,0,0)}{\cal I}_{(0,1,0)}{\cal I}_{h_{12}},
\end{align}
with the hypermultiplet contribution
\begin{align}
{\cal I}_{h_{12}}
&=\oint_C\frac{dz}{2\pi iz}\Pexp(f_h^{12}(z+z^{-1})).
\label{iintersection}
\end{align}
Let us look at the pole structure of the integrand.

\begin{figure}[htb]
\centering
\includegraphics{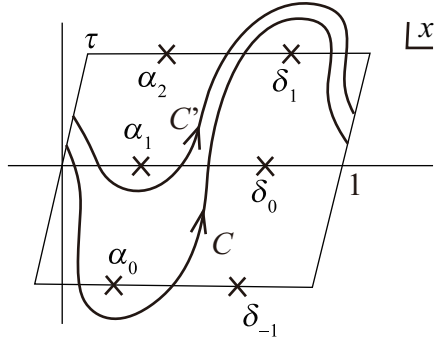}
\caption{The pole structure of the function $\Theta$ on the $x$-plane.}\label{hypx.eps}
\end{figure}

The $q$-expansion of the hypermultiplet single-particle index (\ref{is12})
is given by
\begin{align}
f_h^{12}
&=
\lim_{M\rightarrow\infty}
\left(\sum_{n=0}^M z_{\alpha_n}
-\sum_{n=1}^M (z_{\beta_n}+z_{\gamma_n})
+\sum_{n=2}^M z_{\delta_n}\right)
\label{ihyp}
\end{align}
where $z_{\alpha_n}$, $z_{\delta_n}$, $z_{\beta_n}$, and $z_{\gamma_n}$ are
positions of poles $\alpha_n$, $\delta_n$ and zeros $\beta_n$, $\gamma_n$ of the integrand.
They are given by
\begin{align}
z_{\alpha_n}&=q^{-1}u_3^{\frac{1}{2}}(qu_3)^{n},\nonumber\\
z_{\beta_n}&=q^{-\frac{1}{2}}yu_3^{-\frac{1}{2}}(qu_3)^n,\nonumber\\
z_{\gamma_n}&=q^{-\frac{1}{2}}y^{-1}u_3^{-\frac{1}{2}}(qu_3)^n,\nonumber\\
z_{\delta_n}&=u_3^{-\frac{3}{2}}(qu_3)^n,
\label{abcdn}
\end{align}
and satisfy the relations
\begin{align}
z_{\gamma_n}=z_{\beta_{1-n}}^{-1},\quad
z_{\delta_n}=z_{\alpha_{1-n}}^{-1},\quad
z_{\alpha_n}z_{\delta_n}=z_{\beta_n}z_{\gamma_n}.
\end{align}
The plethystic exponential of $f_h^{12}(z+z^{-1})$ is given by
\begin{align}
\Pexp(f_h^{12}(z+z^{-1}))
=\frac{q^2}{u_3}\Theta(z),
\label{pexpih}
\end{align}
where the function $\Theta(z)$ is defined by
\footnote{In the Schur limit
 the zeros and poles
of the integrand of the hypermultiplet contribution
(\ref{iintersection})
collide:
$\beta_n\rightarrow\alpha_n$,
$\gamma_n\rightarrow \delta_n$,
and the elliptic function becomes trivial: $\Theta_{12}(z)=1$.}
\begin{align}
\Theta(z)&=
\lim_{M\rightarrow\infty}
\prod_{n=1-M}^M
\frac{
(z-z_{\beta_n})(z-z_{\gamma_n})
}
{
(z-z_{\alpha_n})(z-z_{\delta_n})
}.
\label{pexpih2}
\end{align}
This function satisfies
\begin{align}
\Theta(z)=\Theta(qu_3z)=\Theta(z^{-1}).
\label{thetaprop}
\end{align}
Let us define $x$ and $\tau$ by $z=e^{2\pi ix}$ and $qu_3=e^{2\pi i\tau}$, respectively.
If we regard $\Theta$ as the function of $x$,
$\Theta(x)$ is an elliptic function with periods $1$ and $\tau$,
and satisfies $\Theta(x)=\Theta(-x)$.
The position of poles on the $z$-plane are
$z=z_{\alpha_n}$ and $z=z_{\delta_n}$.
They appear on the $x$-plane as two poles
in the fundamental region.
(Figure \ref{hypx.eps})
Due to (\ref{thetaprop}) all residues are the same
up to the signs:
\begin{align}
\Res_{z\rightarrow z_{\alpha_n}}\frac{q^2\Theta}{zu_3}
=-\Res_{z\rightarrow z_{\delta_n}}\frac{q^2\Theta}{zu_3}
=\frac{1}{1-u_3^3}\left(
\frac{1}{u_3}q^2
-\frac{(1+u_3^3)(y+y^{-1})}{u_3^2}q^{\frac{5}{2}}
+\cdots
\right)
\label{adres}
\end{align}

Now, let us calculate the contour integral.
The positive poles read off from the single-particle index
(\ref{ihyp}) are
\begin{align}
z=z_{\alpha_n}\quad(n=0,1,2,\ldots),\quad
z=z_{\delta_n}\quad(n=2,3,4,\ldots).
\label{positibvehyo}
\end{align}
The contour on the $z$-plane enclosing poles in (\ref{positibvehyo}) and $z=0$
is the cycle $C$ in the torus shown in Figure \ref{hypx.eps}.
The contour integral gives
\begin{align}
{\cal I}_{h12}\mbox{ with $C$}
=&
\frac{1}{1-u_3^3}\left(
\frac{3}{u_3}q^2
-\frac{(2+4u_3^3)(y+y^{-1})}{u_3^2}q^{\frac{5}{2}}
+\cdots
\right)
\end{align}
Using this, we obtain ${\cal I}_{(1,1,0)}$ with the most singular part
\begin{align}
{\cal I}_{(1,1,0)}\sim
-\frac{1}{\epsilon_3(\epsilon_1-\epsilon_2)^2(\epsilon_2-\epsilon_3)(\epsilon_3-\epsilon_1)}q^{2N+4}.
\label{anpoles1}
\end{align}
Disappointingly, (\ref{anpoles1})
does not cancel the pole at $u_I=1$ of (\ref{wbospole}).
This means that we need to choose another contour.
Because all residues are the same up to sign,
we can change the integral by only a multiple of the residue
(\ref{adres}).
Fortunately, we can find a contour with which the pole cancellation works.
If we exclude one pole at $z=z_{\alpha_0}$
and use the contour $C'$ in Figure \ref{hypx.eps},
the integral becomes
\begin{align}
{\cal I}_{h_{12}}\mbox{ with $C'$}
=\frac{1}{1-u_3^3}\left(
\frac{2}{u_3}q^2
-\frac{(1+3u_3^3)(y+y^{-1})}{u_3^2}q^{\frac{5}{2}}
+\cdots
\right).
\end{align}
The most singular part of ${\cal I}_{(1,1,0)}$ becomes
\begin{align}
{\cal I}_{(1,1,0)}
\sim-\frac{2}{3\epsilon_3(\epsilon_1-\epsilon_2)^2(\epsilon_2-\epsilon_3)(\epsilon_3-\epsilon_1)}q^{2N+4}.
\label{anpoles}
\end{align}
and this successfully cancel
not only most singular part shown in
(\ref{anpoles}) but also all other divergences.

How should we treat the adoption of the contour $C'$,
which is different from $C$ determined by the pole selection rule?
It is important that the presence of two kings of poles,
positive and negative ones, is related not to the function
$\Theta(z)$ itself but to the single-particle index.
There may be the case that two different single-particle indices
give the same plethystic exponential up to unimportant factors,
and then the definition of the positive and negative poles
may depend on which single-particle index is adopted.
In fact, we can find a single-particle index with which
the pole selection rule gives the desired contour $C'$.
We can rewrite (\ref{hyper1}) as
\begin{align}
(\mbox{hyper})
=\prod_{I=1}^3
\prod_{a=1}^{n_I}
\prod_{b=1}^{n_{I+1}}
\frac{q^{\frac{3}{2}}}{y}\Pexp
f'^{I,I+1}_h\left(a'_{I,I+1}\frac{z_{I,a}}{z_{I+1,b}}
-a'^{-1}_{I,I+1}\frac{z_{I+1,b}}{z_{I,a}}\right),
\label{hyper2}
\end{align}
where $a'_{I,I+1}=(qu_3)^{-\frac{1}{2}}a_{I,I+1}$ and
we defined the modified single-particle index
\begin{align}
f'^{12}_h
=\frac{q^{-\frac{1}{2}}u_3(1-q^{\frac{1}{2}}yu_3^{-1})(1-q^{\frac{3}{2}}y^{-1})}{1-qu_3}.
\label{iprimeh}
\end{align}
We can show that (\ref{hyper2}) is the same as (\ref{hyper1})
by using the relation
\begin{align}
f'^{12}_h
=(qu_3)^{\frac{1}{2}}f_h^{12}-y+q^{\frac{3}{2}}=(qu_3)^{-\frac{1}{2}}f_h^{12}+y^{-1}-q^{-\frac{3}{2}}.
\end{align}
Therefore, this does not affect the integrand in (\ref{in1n2n3}).
However, the difference of the single-particle indices
affects the definition of the positive and negative poles,
and the rule with $f_h'^{12}$ gives the contour $C'$.
With the modified single-particle index
we can keep the pole selection rule intact.
We will use the modified single-particle index for all calculations in the following.

Finally, let us compare
(\ref{sixterms})
and (\ref{singleerror}).
The total double-wrapping contribution (\ref{sixterms})
for small $N$
calculated with the pole selection rule described above are
as follows:
\begin{align}
{\cal I}^{(2)}_{U(1)}&\eqo 112q^6-516q^{\frac{13}{2}}+1869q^7-5394q^{\frac{15}{2}}+14010q^8-32850q^{\frac{17}{2}}+\cdots,
\nonumber\\
{\cal I}^{(2)}_{U(2)}&\eqo 252q^8-1218q^{\frac{17}{2}}+4477q^9-13164q^{\frac{19}{2}}+34452q^{10}+\cdots,
\nonumber\\
{\cal I}^{(2)}_{U(3)}&\eqo 504q^{10}-2548q^{\frac{21}{2}}+9549q^{11}-28734q^{\frac{23}{2}}+76506q^{12}+\cdots,
\nonumber\\
{\cal I}^{(2)}_{U(4)}&\eqo 924q^{12}-4860q^{\frac{25}{2}}+18573q^{13}-57170q^{\frac{27}{2}}+155256q^{14}+\cdots.
\end{align}
These agree with many terms appearing in (\ref{singleerror}).

The error terms are expected to be of order $q^{3N+9}$.
However, it is difficult to explicitly calculate them for $N\geq2$
due to the limited computational resources.
Not only the physical range $N\geq1$,
we can also use unphysical values $N=0$ and $N=-1$ for consistency check.
$N=0$ is the trivial theory with ${\cal I}_{U(0)}=1$ and
$N=-1$ is the empty theory with ${\cal I}_{U(-1)}=0$.
For $N=-1,0,1$, the errors $\Delta_{U(N)}^{(3)}={\cal I}_{U(N)}^{\rm gauge}-({\cal I}_{\rm KK}+{\cal I}^{(1)}+{\cal I}^{(2)})$ are given by
\begin{align}
\Delta{\cal I}^{(3)}_{U(-1)}&\eqo -135q^6-810q^{\frac{13}{2}}
+19977q^7-175796q^{\frac{15}{2}}+1075707q^8+\cdots,\nonumber\\
\Delta{\cal I}^{(3)}_{U(0)}&\eqo -495q^9-2970q^{\frac{19}{2}}+70512q^{10}+\cdots,\nonumber\\
\Delta{\cal I}^{(3)}_{U(1)}&\eqo -1485q^{12}-8910q^{\frac{25}{2}}+211032q^{13}+\cdots
\label{errors3}
\end{align}
and the leading terms are of order $q^{3N+9}$ as expected.

%%%%%%%%%%%%%%%%%%%%%%%%%%%%%%%%%%%%%%%%%%%%%%%%%%
\subsection{$n=3$}
At $n=3$ there are $10$ brane configurations.
\begin{align}
{\cal I}_{U(N)}^{(3)}
=&{\cal I}_{\rm KK}
\big(
{\cal I}_{(3,0,0)}+
{\cal I}_{(0,3,0)}+
{\cal I}_{(0,0,3)}+
{\cal I}_{(2,1,0)}+
{\cal I}_{(0,2,1)}+
{\cal I}_{(1,0,2)}
\nonumber\\&
+{\cal I}_{(1,2,0)}+
{\cal I}_{(0,1,2)}+
{\cal I}_{(2,0,1)}
+{\cal I}_{(1,1,1)}
\big).
\label{tenterms}
\end{align}
For $9$ of them we can calculate the contour integral
with the omission of $a'_{I,J}$.
With the pole selection rule,
${\cal I}_{(3,0,0)}$ is given by
\begin{align}
{\cal I}_{(3,0,0)}&=-\frac{(qu_1)^{3N}\times 3u_1^{29}}
{
\left(
\begin{array}{r}
(u_1-u_2)
(u_1^2-u_2^2)
(u_1^3-u_2^3)
(u_1^4-u_2)\\
\times
(u_1-u_3)
(u_1^2-u_3^2)
(u_1^3-u_3^3)
(u_1^4-u_3)
\end{array}
\right)
}q^9
\nonumber\\
&+
\frac{(qu_1)^{3N}\times 3 u_1^{29} (u_1^4 u_3+ u_1^4 u_2- u_2^2-u_3^2)}
{
\left(
\begin{array}{l}
(1-u_1^3)
(u_1-u_2)
(u_1^2-u_2^2)
(u_1^3-u_2^3)
(u_1^4-u_2)\\
\times
(u_1-u_3)
(u_1^2-u_3^2)
(u_1^3-u_3^3)
(u_1^4-u_3)
\end{array}
\right)
}(y+y^{-1})q^{\frac{19}{2}}+\cdots,
\end{align}
and ${\cal I}_{(0,3,0)}$ and ${\cal I}_{(0,0,3)}$ are obtained by the permutations
among $u_I$.
To calculate ${\cal I}_{(2,1,0)}$
we use the pole selection rule with the modified hypermultiplet single-particle index.
The result is
\begin{align}
{\cal I}_{(2,1,0)}&=
-\frac{(qu_1)^{2N}(qu_2)^N\times 3u_1^{15} u_2^5}{
\left(\begin{array}{c}
(1-u_1^3)
(1-u_3^3)
(u_1-u_2)^2\\
\times
(u_1^2-u_2^2)
(u_1^2-u_3^2)
(u_2-u_3)
(u_1-u_3^4)
\end{array}\right)
}q^9
\nonumber\\
&+\frac{(qu_1)^{2N}(qu_2)^N\times 3 u_1^{12} u_2^2(1 - u_1^3 u_2^3 + u_1^2 u_2^4 - u_1^5 u_2^4)}
{
\left(\begin{array}{c}
(1-u_1^3) (u_1-u_2)^2 (u_1^2-u_2^2) (u_1-u_3)\\
\times
(u_1^2-u_3^2) (u_2-u_3) (1-u_3^3) (u_1-u_3^4)
\end{array}\right)
}(y+y^{-1})q^{\frac{19}{2}}+\cdots,
\label{triples1}
\end{align}
and other five contributions are obtained by the permutations among $u_I$.
We need a special care about ${\cal I}_{(1,1,1)}$.
In this case the quiver diagram has a loop,
and we cannot simply neglect the fugacities $a'_{I,I+1}$.
Unlike the other nine contributions ${\cal I}_{(1,1,1)}$ does not have
$q^{3N+9}$ term and starts at $q^{3N+\frac{19}{2}}$:
\begin{align}
{\cal I}_{(1,1,1)}=
\frac{ (u_1^2+u_2^2+u_3^2-u_1u_2-u_2u_3-u_3u_1) (a'^{-1}_{\rm loop} +a'_{\rm loop}  - 4 (y^{-1} + y))}
{(1-u_1^3)(1-u_2^3)(1-u_3^3)
(u_1-u_2)^2(u_2-u_3)^2(u_3-u_1)^2}
q^{3N+\frac{19}{2}}+\cdots,
\end{align}
where $a'_{\rm loop}\equiv a'_{12}a'_{23}a'_{31}$.

Concerning ${\cal O}(q^{3N+9})$ terms,
the poles at $u_I=0$ successfully cancel among the nine contributions.
For the higher order terms we need to include ${\cal I}_{(1,1,1)}$,
which depends on $a'_{\rm loop}$.
If we require the cancellation among poles at $u_I=1$
we need to set\footnote{Although we introduced
three fugacities $a'_{12}$, $a'_{23}$, and $a'_{31}$
to make the cyclic symmetry manifest,
the results depends only on the product $a'_{\rm loop}$,
and we can practically set $(a'_{12},a'_{23},a'_{31})=(1,1,y^{\pm1})$.}
\begin{align}
a'_{\rm loop}=y^{\pm1}.
\label{aprime}
\end{align}
The result up to the order we have calculated
is symmetric under $a\rightarrow a^{-1}$ and we cannot choose one of $y$ and $y^{-1}$.

After substituting
(\ref{aprime}),
we obtain the following results.
\begin{align}
{\cal I}_{U(-1)}^{(3)}
\eqo &-135q^6-810q^{\frac{13}{2}}+19977q^7+\cdots,\nonumber\\
{\cal I}_{U(0)}^{(3)}
\eqo &-495q^9-2970q^{\frac{19}{2}}+70512q^{10}+\cdots,\nonumber\\
{\cal I}_{U(1)}^{(3)}
\eqo &-1485q^{12}-8910q^{\frac{25}{2}}+211032q^{13}+\cdots.
\end{align}
These agree with the terms shown in (\ref{errors3}).

%%%%%%%%%%%%%%%%%%%%%%%%%%%%%%%%%%%%%%%%%%%%%%%%%%%%%%%%%%%%%%%%%%%%%
\subsection{$n\geq4$}
\begin{figure}[htb]
\centering
\includegraphics{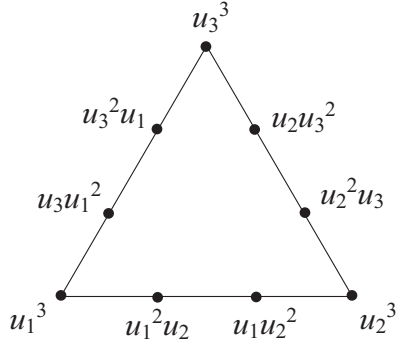}
\caption{Terms in the polynomial $\Delta_n$ for $n=3$ are shown.}\label{triangle}
\end{figure}

For $n\geq4$ the computational cost rapidly increases
and we can say only little about these contributions
due to the limited computational resources.
At $n=4$ there are four essentially different contributions:
${\cal I}_{(4,0,0)}$,
${\cal I}_{(3,1,0)}$,
${\cal I}_{(2,2,0)}$, and
${\cal I}_{(2,1,1)}$,
and other contributions are obtained from these by simple permutations
among $u_I$.
We obtained the leading terms of some of them,
which is of order $q^{4N+16}$.
From these results and
the results for $n=1,2,3$
we can guess the general form of the $q^{nN+n^2}$ term
in ${\cal I}_{(n_1,n_2,n_3)}$ as follows:
\begin{align}
{\cal I}_{(n_1,n_2,n_3)}|_{q^{nN+n^2}}
={\cal I}_{\rm cl}\times n(-q)^{n^2}u_1^{nn_1}u_2^{nn_2}u_3^{nn_3}
\Pexp\left(\frac{\Delta_n}{u_1^{n_1}u_2^{n_2}u_3^{n_3}}-1\right),
\label{leading}
\end{align}
where $\Delta_n$ are the following polynomials of $u_I$.
\begin{align}
\Delta_n=\sum_{k=0}^{n-1}(u_1^ku_2^{n-k}+u_2^ku_3^{n-k}+u_3^ku_1^{n-k}).
\end{align}
By definition monomials of $u_I$ are associated with the points
in the $SU(3)$ weight lattice, and with this relation
the $3n$ terms in $\Delta_n$ form a triangle in the lattice
(Figure \ref{triangle}).
If all $n_I$ are non-zero (\ref{leading}) vanishes.

Although we have not yet confirmed that
(\ref{leading}) reproduces the correct index of the gauge theory,
it is easy to check that it satisfies the pole cancellation condition
for different values of $n$,
and sums up to
\begin{align}
{\cal I}_{U(N)}^{(n)}
\eqo (-1)^n\frac{n(N+4n-1)!}{(3n-1)!(N+n)!}q^{nN+n^2}+\cdots.
\end{align}

%%%%%%%%%%%%%%%%%%%%%%%%%%%%%%%%%%%%%%%%%%%%%%%%
\section{Summary and Discussion}\label{disc.sec}
We proposed a prescription to calculate the superconformal index
of ${\cal N}=4$ $U(N)$ SYM
up to an arbitrary order of the $q$-expansion
based on the idea of \cite{Arai:2019xmp} and
the prescription for the Schur index in \cite{Arai:2020qaj}.
We fixed the selection rule for poles associated with hypermultiplets
and $U(1)_{\rm loop}$ fugacity $a'_{\rm loop}$
so that the poles
at $u_I\rightarrow 1$ cancel, and confirmed that the index obtained with the
prescription agrees with the correct one as far as
we have checked numerically.

Our formula (\ref{theformula}) is based on some assumptions,
and at this moment we do not have any proof.
It is desirable to check whether it works for $n\geq4$.
This requires effective method to carry out the contour integrals.
It would be also important to test our prescription
in other systems than the ${\cal N}=4$ SYM.
The single-wrapping contributions were analyzed
for 4d orbifold quiver gauge theories \cite{Arai:2019wgv},
4d toric quiver gauge theories \cite{Arai:2019aou},
6d ${\cal N}=(2,0)$ theories \cite{Arai:2020uwd},
and 6d ${\cal N}=(1,0)$ theories \cite{Fujiwara:2021xgu}.
In \cite{Arai:2020uwd} multiple-wrapping contributions to the 6d Schur-like index
were partially calculated and the agreement with the results in
\cite{Kim:2013nva,Beem:2014kka} was found.
More detailed analysis of these and other systems including multiple wrapping contributions
are desired.

\begin{figure}[htb]
\centering
\includegraphics{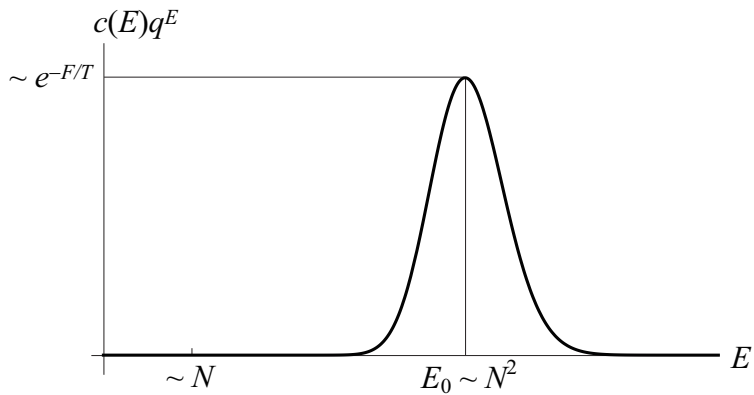}
\caption{Rough behavior of the $q$-expansion of the partition function in the thermodynamic limit.}\label{plot.eps}
\end{figure}

It was recently found that the superconformal index has the information
about the entropy of the AdS blackhole \cite{Choi:2018hmj}.
In general, in the statistical derivation of
thermodynamic quantities
we first calculate the partition function
$Z=\sum_kc_{E_k}q^{E_k}$,
where $E_k$ is energy levels, $c_{E_k}$ is the degeneracies
of each level, and $q=e^{-1/T}$ is the damping factor of the canonical
ansamble.
In the thermodynamic situation we can regard $c_{E_k}$ as
a smooth function $c(E)$ of the energy $E$, and
the summand $c(E)q^E$ of the partition function
has a sharp peak at $E=E_0$
(Figure \ref{plot.eps}).
From the summand $c(E)q^E$ and the coefficient $c(E)$ at the peak
we can determine the free energy $F=-T\log (c(E_0)q^{E_0})$ and
the entropy $S=\log c(E_0)$.
The analysis in \cite{Choi:2018hmj} revealed that
even if we replace the partition function
by the superconformal index,
we can still obtain the entropy by tuning fugacity variables appropriately.
When we discuss macroscopic black hole
we usually consider parameter region with
$E\sim{\cal O}(N^2)$, while the
focus in this paper is
the region $E\sim{\cal O}(N^1)$.
It is important to clarify how these two regions are interpolated.
If our prescription works for
arbitrarily large $n$,
it may be possible
to reproduce the blackhole entropy on the AdS side
without using the quantum gravity.

\section*{Acknowledgments}
The author would like to thank S.~Fujiwara for providing some results
of numerical analysis in Section \ref{s5.sec}.
The author also thank R.~Arai for collaboration at the early stage of this work.
The work of Y.~I. was
partially supported by Grand-in-Aid for Scientific Research (C) (No.21K03569),
Ministry of Education, Science and Culture, Japan.
This work used computational resources TSUBAME3.0 supercomputer provided by Tokyo Institute of Technology.

%%%%%%%%%%%%%%%%%%%%%%%%%%%%%%%%%%%%%%%%%%%%%%%%%%%%%%%%%%%%%%%%%%%%
\appendix
%%%%%%%%%%%%%%%%%%%%%%%%%%%%%%%%%%%%%%%%%%%%%%%%%%%%%%%%%%%%%%%%
\section{Single-particle indices}\label{spi.app}
\subsection{Modes on a wrapped D3-brane}\label{d33modes}
The single particle indices $f_v^I$
were first derived in \cite{Arai:2019xmp}
by using the automorphism between
the supersymmetry algebra on the boundary and that on the wrapped D3-brane.
In this appendix we derive the same result
more directly.

A half of the $32$ supersymmetries is broken by the introduction of
a wrapped D3-brane and modes on the D3-brane form a representation of
the preserved algebra.
Let us start with specifying this algebra.

We express $\bm{S}^5$ as the subset of $\CC^3$ defined by
$|z_1|^2+|z_2|^2+|z_3|^2=1$ where $z_1,z_2,z_3\in\CC$ are three complex coordinates in $\CC^3$.
We define three Cartan generators $R_1$, $R_2$, and $R_3$ acting on three complex coordinates.
We introduce a D3-brane wrapped on the large $\bm{S}^3$ defined by $z_3=0$,
and this breaks $SU(4)_R$ acting on $\bm{S}^5$ to
$SO(4)_{z_1,z_2}\times SO(2)_{z_3}=SU(2)_R\times SU(2)_{\ol R}\times U(1)_{R_3}$,
where the Cartans of the two $SU(2)$ factors are
\begin{align}
R=\frac{R_1-R_2}{2},\quad
\ol R=\frac{R_1+R_2}{2}.
\end{align}

The supercharges $Q$, $S$, $\ol Q$, and $\ol S$ belong to quartet (the fundamental or the anti-fundamental) representations
of $SU(4)_R$.
Each of them is decomposed into a pair of doublets of the unbroken symmetry $SU(2)_R\times SU(2)_{\ol R}\times U(1)_{R_3}$.
The one from each pair satisfying $H=R_3$ is preserved by the D3-brane and the other is broken.
The preserved supersymmetries are summarized in Table \ref{d3q.tbl}.
\begin{table}[htb]
\caption{The quantum numbers of the preserved supercharges on a wrapped D3-brane are shown.
$[\frac{1}{2}]$ represents $SU(2)$ doublet.}\label{d3q.tbl}
\centering
\begin{tabular}{cccccccc}
\hline
\hline
& $H$ & $J$ & $\ol J$ & $R$ & $\ol R$ & $R_3$ & $Y$ \\
\hline
$Q^A_a$ & $+\frac{1}{2}$ & $[\frac{1}{2}]$ & $0$ & $[\frac{1}{2}]$ & $0$ & $+\frac{1}{2}$ & $+1$ \\
$S_A^a$ & $-\frac{1}{2}$ & $[\frac{1}{2}]$ & $0$ & $[\frac{1}{2}]$ & $0$ & $-\frac{1}{2}$ & $-1$ \\
$\ol Q_{\dot A}^{\dot a}$ & $+\frac{1}{2}$ & $0$ & $[\frac{1}{2}]$ & $0$ & $[\frac{1}{2}]$ & $+\frac{1}{2}$ & $-1$ \\
$\ol S^{\dot A}_{\dot a}$ & $-\frac{1}{2}$ & $0$ & $[\frac{1}{2}]$ & $0$ & $[\frac{1}{2}]$ & $-\frac{1}{2}$ & $+1$ \\
\hline
\end{tabular}
\end{table}
$Y$ is the generator of the outer automorphism $U(1)_Y$ of $psu(2,2|4)$ rotating the supercharges by $Q\rightarrow e^{i\alpha}Q$.
Note that the preserved components of $Q$ and $\ol Q$ anti-commute among them, and
the algebra does not include the $K$ and $P$ generators.
The algebra generated by these supercharges is
the product ${\cal A}_L\times{\cal A}_R$ of two isomorphic algebras
${\cal A}_L\approx{\cal A}_R\approx su(2|2)$, which are generated by
the following generators:\footnote{$R=R^1{}_1$ and $\ol R=\ol R^{\dot1}{}_{\dot1}$}
\begin{align}
{\cal A}_L&:\quad
Q^A_a,\quad
S_A^a,\quad
Z,\quad
J^a{}_b,\quad
R^A{}_B,\nonumber\\
{\cal A}_R&:\quad
\ol Q_{\dot A}^{\dot a},\quad
\ol S^{\dot A}_{\dot a},\quad
\ol Z,\quad
\ol J^{\dot a}{}_{\dot b},\quad
\ol R^{\dot A}{}_{\dot B}.
\end{align}
$Z$ and $\ol Z$ are central elements and
they are both identified with $H-R_3$.

Important anti-commutation relations for the construction of representations are
\begin{align}
2\{S_A^a,Q^B_b\}&=\delta^a_b\delta_A^BZ+2\delta^B_AJ^a{}_b+2\delta^a_bR^B{}_A,\nonumber\\
2\{\ol Q^{\dot a}_{\dot A},\ol S_{\dot b}^{\dot B}\}&=\delta^{\dot a}_{\dot b}\delta_{\dot A}^{\dot B}\ol Z
-2\delta^{\dot B}_{\dot A}\ol J^{\dot a}{}_{\dot b}-2\delta^{\dot a}_{\dot b}\ol R^{\dot B}{}_{\dot A}.
\label{sqsqac}
\end{align}
Other anti-commutation relations among supercharges vanish.
From these we obtain the bounds
\begin{align}
Z-2|J|-2|R|\geq0,\quad
\ol Z-2|\ol J|-2|\ol R|\geq0.
\label{bounds}
\end{align}

\begin{figure}[htb]
\centering
\includegraphics{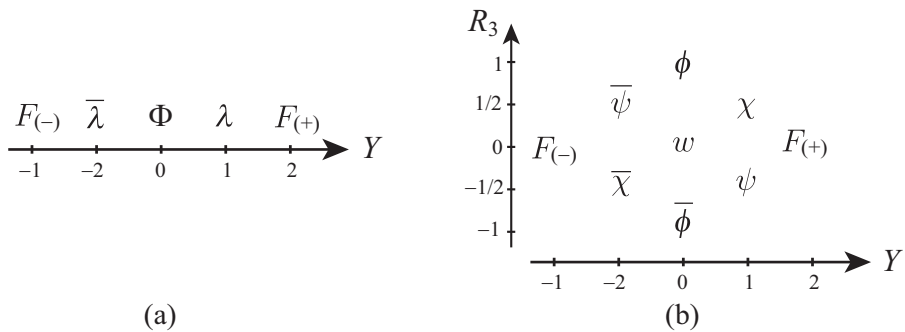}
\caption{The vector multiplet on a D3-brane includes
five components distinguished by $U(1)_Y$ charges
as shown in (a).
Each of them belongs to an irreducible representation
of $SU(4)_R^{\rm D3}$.
They are decomposed into nine irreducible representations of
$SU(2)_J\times SU(2)_{\ol J}\times U(1)_{R_3}$ as shown in (b).}\label{vector.eps}
\end{figure}

Next, let us consider component fields of the vector multiplet on the D3-brane.
A vector multiplet consists of five component fields distinguished by $Y$
(Figure \ref{vector.eps} (a)).
On a flat D3-brane in the flat spacetime they belong to representations
of the R-symmetry $SU(4)_R^{\rm D3}$.
We emphasize that $SU(4)_R^{\rm D3}$ is different from the R-symmetry
of the boundary theory $SU(4)_R$ acting on $\bm{S}^5$.
$SU(4)_R^{\rm D3}$ acts on the six transverse directions of the wrapped D3:
four in $AdS_5$ and two in $\bm{S}^5$.
Due to the curvature of the background spacetime $SU(4)_R^{\rm D3}$ is broken to $SO(4)\times SO(2)=SU(2)_J\times SU(2)_{\ol J}\times U(1)_{R_3}$.
The two $SU(2)$ factors are identified with the spins in AdS.
The component fields are decomposed into nine irreducible representations
of this subgroup as in Figure \ref{vector.eps} (b).

We can determine the Kaluza-Klein spectrum
without using the detailed information of the theory
(equations of motion of the fields)
thanks to the large symmetry.
In general, we can construct an irreducible representation by
acting $Q$ and $\ol Q$ as raising operators on primary states.
Because both $Q$ and $\ol Q$ carry $R_3=+\frac{1}{2}$
we can use $R_3$ as ``the level'' in the construction.
The fields on the D3-brane carry different values on $R_3$
as is shown in Figure \ref{vector.eps} (b),
and $\ol\phi$ carries the smallest value $R_3=-1$.
The scalar field $\ol\phi$ is expanded into $\bm{S}^3$ spherical harmonics
belonging to the $SU(2)_R\times SU(2)_{\ol R}$ representation
\begin{align}
\bigoplus_{\ell=0}^\infty ([\tfrac{\ell}{2}],[\tfrac{\ell}{2}]).
\end{align}
Let $\phi^{(\ell)}$ be the set of modes belonging to $([\tfrac{\ell}{2}],[\tfrac{\ell}{2}])$.
For each $\ell$ we can construct an irreducible representation of ${\cal A}$ by
acting the raising operators $Q$ and $\ol Q$ repeatedly on $\ol\phi^{(\ell)}$.

The diagram in Figure \ref{vector.eps} (b) shows that $Q^3$ and $\ol Q^3$ eliminate
$\ol\phi$ and
the representation must be short
for both ${\cal A}_L$ and ${\cal A}_R$.
Hence the highest weight state $(\tfrac{\ell}{2},\tfrac{\ell}{2})$ in the harmonics
should saturate both the bounds in (\ref{bounds}).
This determines the common central charge
$H-R_3$ as
\begin{align}
H-R_3=\ell.
\end{align}
Now we have completely determined the quantum numbers
of modes of $\ol\phi$.
See the line of $\ol\phi^{(\ell)}$ in Table \ref{d3spec.tbl}.
\begin{table}[htb]
\caption{Fluctuation modes on a wrapped D3-brane.
$[s]$ represents the spin-$s$ $SU(2)$ representation.}\label{d3spec.tbl}
\centering
\begin{tabular}{cccccccc}
\hline
\hline
& $H$ & $J$ & $\ol J$ & $R$ & $\ol R$ & $R_3$ & $Y$ \\
\hline
$\ol\phi^{(\ell)}$ & $\ell-1$           & $0$ & $0$             & $[\frac{\ell}{2}]$ & $[\frac{\ell}{2}]$ & $-1$           & $0$ \\
$\ol\chi^{(\ell)}$ & $\ell-\frac{1}{2}$ & $0$ & $[\frac{1}{2}]$ & $[\frac{\ell}{2}]$ & $[\frac{\ell-1}{2}]$   & $-\frac{1}{2}$ & $-1$ \\
$F_{(-)}^{(\ell)}$ & $\ell$             & $0$ & $0$             & $[\frac{\ell}{2}]$ & $[\frac{\ell-2}{2}]$ & $0$            & $-2$ \\
\hline
$\psi^{(\ell)}$    & $\ell-\frac{1}{2}$ & $[\frac{1}{2}]$ & $0$ & $[\frac{\ell-1}{2}]$ & $[\frac{\ell}{2}]$ & $-\frac{1}{2}$ & $+1$ \\
$w^{(\ell)}$       & $\ell$             & $[\frac{1}{2}]$ & $[\frac{1}{2}]$ & $[\frac{\ell-1}{2}]$ & $[\frac{\ell-1}{2}]$ & $0$ & $0$ \\
$\ol\psi^{(\ell)}$ & $\ell+\frac{1}{2}$ & $[\frac{1}{2}]$ & $0$ & $[\frac{\ell-1}{2}]$ & $[\frac{\ell-2}{2}]$ & $+\frac{1}{2}$ & $-1$ \\
\hline
$F_{(+)}^{(\ell)}$ & $\ell$ & $0$ & $0$ & $[\frac{\ell-2}{2}]$ & $[\frac{\ell}{2}]$ & $0$ & $+2$ \\
$\chi^{(\ell)}$ & $\ell+\frac{1}{2}$ & $0$ & $[\frac{1}{2}]$ & $[\frac{\ell-2}{2}]$ & $[\frac{\ell-1}{2}]$ & $+\frac{1}{2}$ & $+1$ \\
$\phi^{(\ell)}$ & $\ell+1$ & $0$ & $0$ & $[\frac{\ell-2}{2}]$ & $[\frac{\ell-2}{2}]$ & $+1$ & $0$ \\
\hline
\end{tabular}
\end{table}

The full representation of ${\cal A}_L\times{\cal A}_R$ is constructed by acting $Q$ and $\ol Q$ on the primary states
and removing zero norm states.
Because the algebra is factorized, we can consider excitations of ${\cal A}_L$ and ones of ${\cal A}_R$ separately.
Let us focus on the ${\cal A}_L$ part.
The quantum numbers $(J,R)_Z$ of $\ol\phi$ for the bosonic subalgebra
\begin{align}
su(2)_J\times su(2)_R\times u(1)_Z\subset
{\cal A}_L
\end{align}
are $(0,[\tfrac{\ell}{2}])_\ell$.
By acting $Q$ and removing null states
we obtain the irreducible representation
\begin{align}
{\cal R}_\ell=(0,[\tfrac{\ell}{2}])_\ell\oplus
([\tfrac{1}{2}],[\tfrac{\ell-1}{2}])_\ell\oplus
(0,[\tfrac{\ell-2}{2}])_\ell.
\label{rell}
\end{align}
A spin $s$ representation $[s]$ with negative $s$ should be removed.
Namely, for $\ell=0$ and $\ell=1$ the representations are
\begin{align}
{\cal R}_0=(0,0)_0,\quad
{\cal R}_1=(0,[\tfrac{1}{2}])_1\oplus
([\tfrac{1}{2}],0)_1.
\label{r01}
\end{align}
We can also construct the ${\cal A}_R$ representation ${\ol{\cal R}}_\ell$
as a direct sum of irreducible representations of the bosonic subgroup
$su(2)_{\ol J}\times su(2)_{\ol R}\times u(1)_Z$,
which are given in the same way as (\ref{rell}) and (\ref{r01}).
The modes on a wrapped D3-brane belong to the representation
\begin{align}
\bigoplus_{\ell=0}^\infty
{\cal R}_\ell\otimes\ol{\cal R}_\ell.
\label{oprr}
\end{align}
All modes belonging to (\ref{oprr}) are summarized in Table \ref{d3spec.tbl}.
Using the information in the table
it is straightforward to obtain the single-particle index
\begin{align}
f_v^3=1-\frac{(1-\frac{1}{qu_3})(1-yq^{\frac{3}{2}})(1-y^{-1}q^{\frac{3}{2}})}{(1-qu_1)(1-qu_2)}.
\label{vectorindexond3}
\end{align}

Some comments are in order.

The $\ell=0$ mode of $\phi$,
the unique component of the representation ${\cal R}_0\times\ol{\cal R}_0$
has negative energy.
This does not contradict the supersymmetry
because $H$ always appears in the unbroken algebra
in the form $H-R_3$,
and $H$ itself is not subject to any
bounds.

The $\ell=1$ modes are Nambu-Goldstone modes
associated with the symmetry breaking.
The broken generators
are realized non-linearly on the
D3-brane and behave like creation and
annihilation operators.
$\ell=1$ modes are generated by broken generators,
and the quantum numbers agree with those of the
corresponding generators.
Modes in $\phi^{(1)}$ have $E=0$
and generate degenerate states belonging to the
$SU(4)_R$ representations.
A mode in $\ol\chi^{(1)}$ or $\psi^{(1)}$ saturate
the BPS bound associated with the broken supercharge
which carries the same quantum numbers with the mode.
The bose-fermi pairing caused by such a mode is
the reason for the vanishing of the
index in a special limit like the Schur limit.

%%%%%%%%%%%%%%%%%%%%%%%%%%%%%%%%%%%%%%%%%%%%%%%%%%%%%%%%%%%%%%%%%
\subsection{Modes on the intersection}\label{appa2}
We consider two wrapped D3-branes: one wrapped on $z_1=0$ and
the other wrapped on $z_2=0$.
The supercharges with $R_1=R_2=H$ are preserved.
The quantum numbers of unbroken supersymmetry generators are summarized in Table \ref{intcharge}.
\begin{table}[htb]
\caption{Quantum numbers of the unbroken supersymmetry generators.
$[\frac{1}{2}]$ represents $SU(2)$ doublet.}\label{intcharge}
\centering
\begin{tabular}{c|c|cc|ccc}
\hline
\hline
 & $H$ & $J$ & $\ol J$ & $R_1$ & $R_2$ & $R_3$ \\
\hline
    $Q_a$ & $+\frac{1}{2}$ & $[\frac{1}{2}]$ & $0$ & $+\frac{1}{2}$ & $+\frac{1}{2}$ & $-\frac{1}{2}$ \\
    $S^a$ &$-\frac{1}{2}$ & $[\frac{1}{2}]$ & $0$ & $-\frac{1}{2}$ & $-\frac{1}{2}$ & $+\frac{1}{2}$ \\
$\ol Q^{\dot a}$ & $+\frac{1}{2}$ & $0$ & $[\frac{1}{2}]$ & $+\frac{1}{2}$ & $+\frac{1}{2}$ & $+\frac{1}{2}$ \\
$\ol S_{\dot a}$ &$-\frac{1}{2}$ & $0$ & $[\frac{1}{2}]$ & $-\frac{1}{2}$ & $-\frac{1}{2}$ & $-\frac{1}{2}$ \\
\hline
\end{tabular}
\end{table}
The non-vanishing anti-commutation relations among the preserved supercharges are
\begin{align}
\{S^a,Q_b\}&=\delta^a_b(H-R_1-R_2+R_3)+2J^a{}_b,\nonumber\\
\{\ol Q^{\dot a},\ol S_{\dot b}\}&=\delta^{\dot a}_{\dot b}(H-R_1-R_2-R_3)-2\ol J^{\dot a}{}_{\dot b}.
\label{hypcomm}
\end{align}
Again, the preserved superconformal algebra factorizes
to two isomorphic algebras ${\cal A}\approx\ol{\cal A}\approx su(2|1)$
generated by the following generators:
\begin{align}
{\cal A}:\quad
Q_a,\quad
S^a,\quad
J^a{}_b,\quad
H-R_1-R_2+R_3,\nonumber\\
\ol{\cal A}:\quad
\ol Q^{\dot a},\quad
\ol S_{\dot a},\quad
\ol J^{\dot a}{}_{\dot b},\quad
H-R_1-R_2-R_3.
\end{align}
From (\ref{hypcomm}) we obtain the bounds
\begin{align}
H-2|J|-R_1-R_2+R_3\geq0,\quad
H-2|\ol J|-R_1-R_2-R_3\geq0.
\label{hypbounds}
\end{align}

A single hypermultiplet lives
on the intersection locus of the two D3-branes.
It consists of a chiral multiplet arising from open strings of one orientation
and another chiral multiplet from open strings of the opposite orientation.
The two chiral multiplets carry opposite gauge charges
and the other quantum numbers are the same.
Let us focus on one of these chiral multiplets.
It includes two bosonic and two fermionic degrees of freedom.

In the case of the flat spacetime background
scalar fields belong to a doublet of the $SO(4)$ rotation of the four DN directions
and fermion fields are singlet under the $SO(4)$.
This is not affected by the background curvature.
Namely, we have two bosonic components with $R_1=R_2=\pm\frac{1}{2}$ and fermionic components
with $R_1=R_2=0$.

\begin{figure}[htb]
\centering
\includegraphics{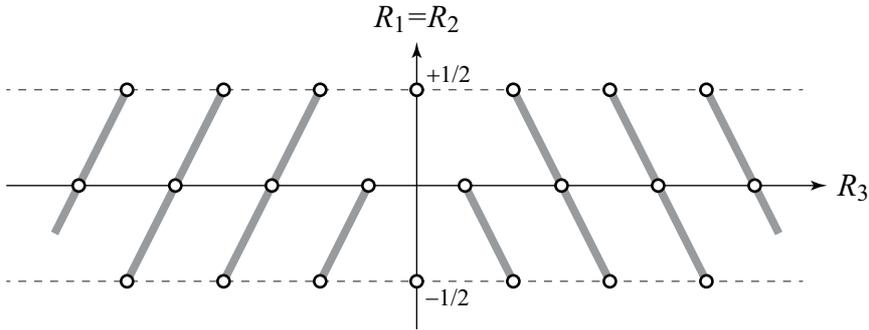}
\caption{The multiplet structure of the modes of
the hypermultiplet on the intersection.
Modes are expressed by circles, and modes in the same
irreducible representation are connected by segments.}\label{hypermodes}
\end{figure}

The charge $R_3$ is the Kaluza-Klein momentum along the intersection.
Because of the non-vanishing $R_3$ of the supercharges
the bosonic and fermionic components satisfy different quantization condition of $R_3$.
Namely we should consider the NS-sector, and $R_3\in\ZZ$ for bosonic components
and $R_3\in\ZZ+1/2$ for fermionic components.
We expand $\wt q$, $q$, and $\psi$ into Fourier modes $\wt q^{(n)}$, $q^{(n)}$, and $\psi^{(r)}$, respectively,
where $n\in\ZZ$ and $r\in\ZZ+\frac{1}{2}$ denote the Kaluza-Klein momentum $R_3$.
By requiring all modes to belong to short superconformal representations
the energy of all modes are determined as shown in Table \ref{hypertbl}.
\begin{table}[htb]
\caption{Quantum numbers of the component fields in a chiral multiplet at the intersection.}\label{hypertbl}
\centering
\begin{tabular}{c|cccc}
\hline
\hline
& $H$ & $(J,\ol J)$ & $R_1=R_2$ & $R_3$ \\
\hline
$q^{(n)}$ & $|n|+1$ & $(0,0)$ & $+\frac{1}{2}$ & $n\in\ZZ$ \\
$\psi_\pm^{(r)}$   & $|r|$ & $\begin{array}{c}([\frac{1}{2}],0)_{r>0}\\(0,[\frac{1}{2}])_{r<0}\end{array}$ & $0$ & $r\in\frac{1}{2}+\ZZ$ \\
$\wt q^{(n)}$ & $|n|-1$ & $(0,0)$ & $-\frac{1}{2}$ & $n\in\ZZ$ \\
\hline
\end{tabular}
\end{table}
The BPS bounds in (\ref{hypbounds}) show that if $n\geq 0$ scalar modes $q^{(n)}$ and $\wt q^{(n)}$ are $\ol{\cal A}$ singlets
and if $n\leq0$ they are ${\cal A}$ singlets.
With this information we can unambiguously determine the multiplet structure as shown in
Figure \ref{hypermodes}.

From the quantum numbers in Table \ref{hypertbl} we can easily calculate
the single particle index:
\begin{align}
f_h^{12}
&=\frac{u_3^{\frac{1}{2}}}{q}
\frac{(1-q^{\frac{3}{2}}y)(1-q^{\frac{3}{2}}y^{-1})}{1-qu_3}.
\label{interi}
\end{align}

%%%%%%%%%%%%%%%%%%%%%%%%%%%%%%%%%%%%%%%%%%%%%%%%%
\section{Mixed poles}\label{mixed.sec}
After repeating the integrals
we may have the following factor in the integrand
at the step of $z_k$-integral.
\begin{align}
\left(1-\frac{f_1f_2\cdots f_p}{f'_1f'_2\cdots f'_q}\frac{z_m}{z_k}\right)^{-1},\quad
p,q\geq1,\quad
p+q\leq k<m.
\label{mixedfactor}
\end{align}
Then, we have a mixed pole at
\begin{align}
z_k=
\frac{f_1f_2\cdots f_p}{f'_1f'_2\cdots f'_q}z_m.
\label{mixedzk}
\end{align}
Although this pole may sit inside the unit circle,
the above pole selection rule does not include mixed poles like this.
In fact, this kind of poles always cancel among them,
and in total do not contribute to the result.
Let us prove this by mathematical induction.

Let us consider poles on the $z_k$-plane
after the integrals with respect to $k-1$ variables
\begin{align}
z_1,\ldots,z_{k-1}.
\label{donez}
\end{align}
We suppose that at every step of these $k-1$ integrals
mixed poles cancel among them,
and based on this assumption
we want to show that this is also the case for
the $z_k$-integral.

Let us first consider how the factor (\ref{mixedfactor}) is
generated by the preceding integrals.
Due to the assumption that
mixed poles cancel in the first $k-1$ integrals,
(\ref{mixedfactor}) should be generated from a factor
\begin{align}
\left(1-f_1\frac{z_{a_1}}{z_c}\right)^{-1}
\label{seed}
\end{align}
by the substitusions
\begin{align}
z_{a_1}=f_2f_3\cdots f_pz_m,\quad
z_c=f'_1f'_2f'_3\cdots f'_qz_k.
\label{substi}
\end{align}
The first substitution in (\ref{substi}) is realized
if the integrand contain
\begin{align}
\left(1-f_2\frac{z_{a_2}}{z_{a_1}}\right)^{-1}
\left(1-f_3\frac{z_{a_3}}{z_{a_2}}\right)^{-1}
\cdots
\left(1-f_p\frac{z_m}{z_{a_{p-1}}}\right)^{-1}
\label{branch1}
\end{align}
and
picking up positive poles in the integrals with respect to
$p-1$ variables
\begin{align}
z_{a_1},z_{a_2},\ldots,z_{a_{p-1}}.
\label{subset1}
\end{align}
The order of the variables in (\ref{subset1}) has nothing to do with
the integration order.
Similarly, the second in (\ref{substi}) is realized
if the integrand contains
\begin{align}
\left(1-f'_1\frac{z_{a'_1}}{z_{c}}\right)^{-1}
\left(1-f'_2\frac{z_{a'_2}}{z_{a'_1}}\right)^{-1}
\cdots
\left(1-f'_q\frac{z_k}{z_{a'_{q-1}}}\right)^{-1}
\label{branch2}
\end{align}
and picking up positive poles from these factors
in the integrals with respect to $q$ variables
\begin{align}
z_{c},z_{a'_1},z_{a'_2},\ldots,z_{a'_{q-1}}.
\label{subset2}
\end{align}
(\ref{subset1}) and (\ref{subset2}) must be mutually exclusive subsets of
(\ref{donez}).

Let $G(z)$ be the product of all factors in
(\ref{seed}), (\ref{branch1}), and (\ref{branch2}),
and $F(z)$ be the remaining factor in the integrand.
Namely, the integrand before starting the integrals is
$F(z)G(z)$.
After performing integrals for variables
(\ref{donez}) by picking up the poles we described above
we obtain
\begin{align}
F(
z_{a_i}\rightarrow \ol z_{a_i},
z_{a'_i}\rightarrow \ol z_{a'_i},
z_c\rightarrow f'_1\cdots f'_qz_k)
\left(1-\frac{f_1f_2\cdots f_p}{f'_1f'_2\cdots f'_q}\frac{z_m}{z_k}\right)^{-1}.
\label{ans1}
\end{align}
where $\ol z_{a_i}$ and $\ol z_{a'_i}$ are the values substituted to $z_{a_i}$ and $z_{a'_i}$, respectively,
after the integrals with respect to the variables in (\ref{donez}).
If (\ref{donez}) contains variables that are not contained in either
(\ref{subset1}) or (\ref{subset2}), $F$ in
(\ref{ans1}) should be understood as the function obtained
by the integrals with respect to them.
Now let us perform the $z_k$-integral around the
mixed pole (\ref{mixedzk}).
The result is
\begin{align}
F(
z_{a_i}\rightarrow \ol z_{a_i},
z_{a'_i}\rightarrow \ol z_{a'_i},
z_c\rightarrow f_1\cdots f_pz_m)
\label{ans1a}
\end{align}
What we want to show is there is always another contribution
that cancels this result.

Notice that every integration variable in (\ref{subset1}) or (\ref{subset2})
appears in $G(z)$ twice.
Except for $z_c$, each of them appears once in a numerator and once in a denominator.
Unlike the other variables $z_c$ appears twice in denominators in the two factors,
the seed factor (\ref{seed}) and the first factor in (\ref{branch2}).
Corresponding to these two factors there are two positive poles
\begin{align}
z_c=f_1z_{a_1},\quad
z_c=f_1'z_{a'_1}
\label{twopoles}
\end{align}
on the $z_c$-plane.
In the calculation above we took the latter in
(\ref{twopoles}) in the $z_c$-integral.

Now, let us exchange the roles of the two poles in
(\ref{twopoles}).
Namely, we pick up the formar in the $z_c$-integral.
Then, as the result of $k-1$ integrals with respect to
the variables in (\ref{donez}), we obtain the following factor:
\begin{align}
F(
z_{a_i}\rightarrow \ol z_{a_i},
z_{a'_i}\rightarrow \ol z_{a'_i},
z_c\rightarrow f_1\cdots f_pz_m)
\left(1-\frac{f'_1f'_2\cdots f'_q}{f_1f_2\cdots f_p}\frac{z_k}{z_m}\right)^{-1}
\label{ans2}
\end{align}
instead of (\ref{ans1}).
$\ol z_{a_i}$ and $\ol z_{a'_i}$ in
(\ref{ans2}) are the same as those in
(\ref{ans1}).
This is similar to (\ref{ans1}), and
produces the pole at the same position
(\ref{mixedzk}).
Furthermore, we can easily confirm that
the $z_k$-integral around the pole (\ref{mixedzk})
gives the negative of (\ref{ans1a}),
and the two contributions calcel each other.
In this way, all mixed poles cancel among them and we only need to take account of
positive poles as is claimed in the pole selection rule.

%%%%%%%%%%%%%%%%%%%%%%%%%%%%%%%%%%%%%%%%%%%
\section{Cut-off order}\label{cutoff.app}
The $q$-expansion of the single-particle index includes infinite terms
and in a practical numerical calculation we need to introduce cut-off at some order.
To obtain correct results up to desired order we need to choose the cut-off order
carefully.
To determine the appropriate order of the cut-off
let us consider the effect of the inclusion of a term
of order $q^m$ to the single-particle index.

Let $m_0$ be the order of the leading term in the $q$-expansion of the
final result.
The inclusion of an ${\cal O}(q^m)$ term into the single-particle index
produces the extra factor
\begin{align}
1-{\cal O}(q^m)\frac{z_i}{z_j}
\label{newfactor}
\end{align}
in the integrand.
If we used the unit circle in the every $z_i$-integral,
this would change the integral by the factor $1-{\cal O}(q^m)$,
and the correction to the final result
would be of order $q^{m_0+m}$.
However, we need to consider the case with tachyonic
terms in the single-particle index.
Let us assume that the leading term $f_1$ of the $q$ expansion of the single-particle
index starts at order $f_1\sim{\cal O}(q^{-t})$.
If $t>0$ we cannot use unit circles for the contours.
We need to deform the contours to include all
positive poles and exclude all negative poles.
In the rank $n$ case we need to perform the integrals $n-1$ times,
and the final $z_{n-1}$-integral, the most distant positive pole from the origin
is $z_{n-1}\sim f_1^{n-1}\sim{\cal O}(q^{-(n-1)t})$,
while the closest negative pole to the origin
is $z_{n-1}\sim f_1^{-(n-1)}\sim{\cal O}(q^{(n-1)t})$.
Therefore, $|z_{n-1}|$ varies on the contour in the following range:
\begin{align}
{\cal O}(q^{(n-1)t})\leq z_{n-1}\leq{\cal O}(q^{-(n-1)t}).
\end{align}
As the result, the factor
(\ref{newfactor}) becomes
\begin{align}
1-{\cal O}(q^{m-(n-1)t}).
\end{align}
Therefore, the inclusion of ${\cal O}(q^m)$
term in the single-particle index affect the result by terms of order
$q^{m_0+m-(n-1)t}$.
Contrary, if we want to obtain the correct answer up to $q^{m_0+c}$ terms,
then it is sufficient if we include terms
up to $q^{m_{\rm max}}$
in the single-particle index, where
\begin{align}
m_{\rm max}=c+(n-1)t.
\end{align}

\end{document}